\documentclass[twocolumn,showpacs,pra,aps,superscriptaddress,floatfix]{revtex4-1}
\usepackage{amsfonts}
\usepackage{mathrsfs}
\usepackage{amssymb}
\usepackage{bm}
\usepackage{graphicx}
\usepackage[colorlinks=true,linkcolor=blue,citecolor=red,urlcolor=blue]{hyperref}

\begin{document}

\title
{Time evolution of decay for purely absorptive potentials:\\ The effect of spectral singularities}
\author{Gast\'on Garc\'{\i}a-Calder\'on}
\email{gaston@fisica.unam.mx}
\affiliation{Instituto de F\'{\i}sica, Universidad Nacional Aut\'onoma de
M\'exico, Apartado Postal {20 364}, 01000 M\'exico, Distrito Federal, M\'exico}
\author{Lorea Chaos-Cador}
\email{lorea@ciencias.unam.mx}
\affiliation{Universidad Aut\'onoma de la Ciudad de M\'exico,
Prolongaci\'on San Isidro 151, 09790 M\'exico, Distrito Federal, M\'exico}

\date{\today}

\begin{abstract}
We consider an analytical approach that involves the complex poles of the propagator to investigate the effect due to the lack of time reversal invariance on the time evolution of decay for purely absorptive potentials. We find that the choice of the initial state may change the exponential decaying regime into a nonexponential oscillatory behavior at time scales of the order of a lifetime of the system.
We illustrate this effect for a spectral singularity corresponding to a purely imaginary delta-shell potential.
\end{abstract}
\pacs{03.65.Ca,03.65.Xp,03.65.Db}

\maketitle

\section{Introduction}\label{uno}
One finds in the literature a great deal of work dealing with complex absorbing potentials as discussed in a recent review by Muga et. al. \cite{muga04}.  Of particular interest in studies involving these potentials are spectral singularities which correspond to points in the continuous spectra of quantum systems where a continuous solution to the Schr\"odiner equation of the problem becomes singular \cite{mostafazadeh09a}. They were mainly studied by mathematicians in the 50s and 60s of last century \cite{naimark54,kemp58,schwartz60} but in recent years they have attracted the attention  of investigations on the properties of $\mathcal{PT}$-symmetric and non-$\mathcal{PT}$-symmetric scattering potentials \cite{samsonov05,mostafazadeh09a,mostafazadeh09b}, and more recently, in laser optics \cite{stone10,longhi10,stone11,mostafazadeh11,longhi11,mostafazadeh13a}. A common feature of these recent works is that they consider systems which are open and hence may be described by imposing purely outgoing boundary conditions on the corresponding wave equations. In the above context we have recently extended the formalism of resonant states to complex potentials \cite{chgc13}.

The purpose of this work is to investigate the effect of absorptive potentials, i.e., purely imaginary potentials, on the time evolution of quantum decay. We find a general condition on the initial state that leads to a nonexponential oscillatory behavior of the survival probability at time scales of the order of the lifetime of the system. Our findings are illustrated by considering an exactly solvable model.

In section \ref{dos}, we review  the formalism of resonant states for complex imaginary potentials and provide the corresponding expressions for the decaying wave function and  the survival probability.  Section \ref{tres} applies the formalism to an attractive purely imaginary $\delta$-shell potential to calculate the survival probability. Finally, Section \ref{cuatro} provides some concluding remarks.

\section{Formalism}\label{dos}

\begin{figure}[!tbp]
\begin{center}
\includegraphics[width=3.5in]{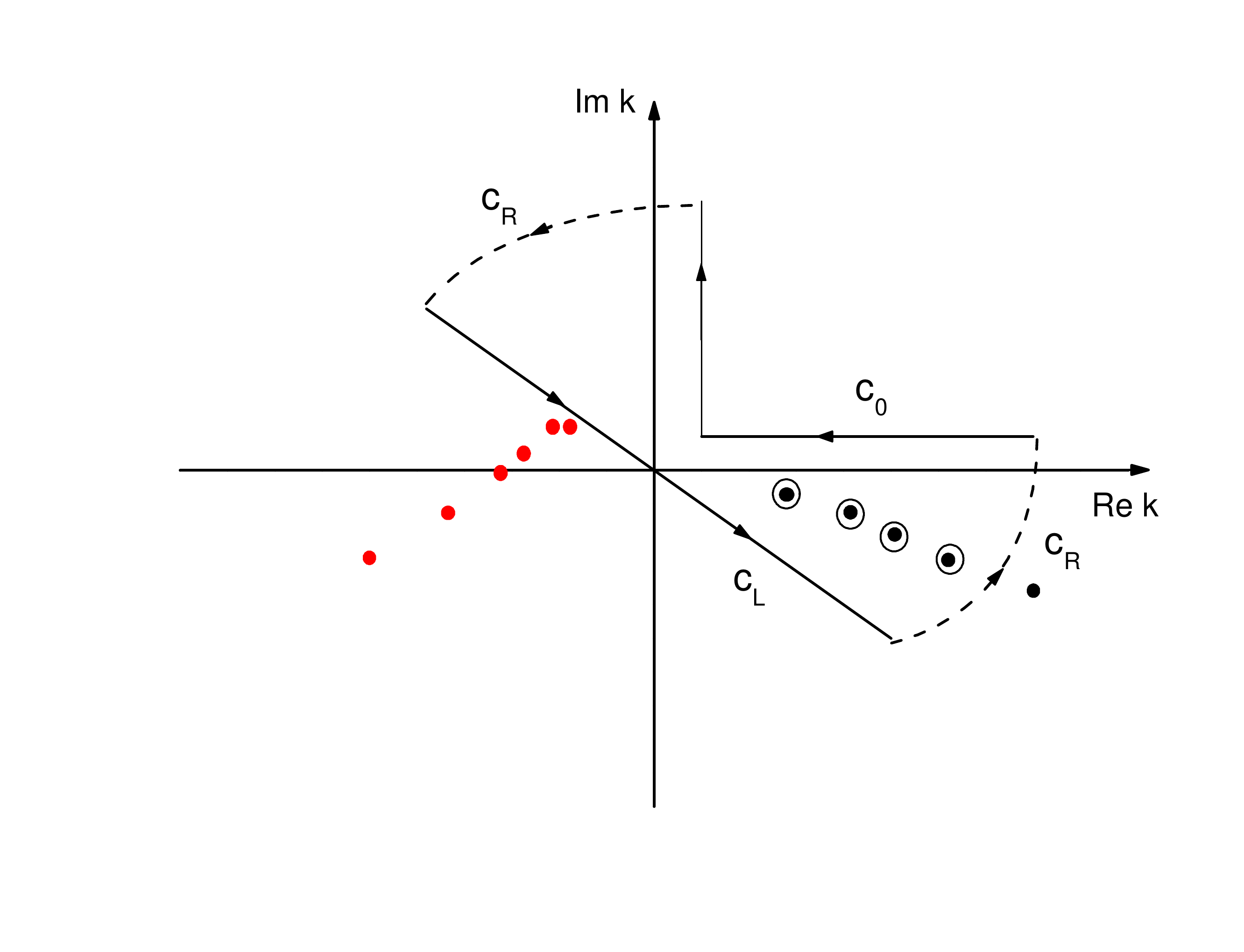}
\end{center}
\caption{\label{fig1} Deformation of the contour $C_0$ in the complex $k$ plane. See text.}
\end{figure}

Let us consider the time evolution of  decay of an initial  wave function  $\psi(r,0)$  confined initially, at $t=0$, along the internal region of a spherically symmetric absorptive potential of finite range, namely,
\begin{equation}
V(r)=-iW(r)
\label{ia1}
\end{equation}
with $W(r)$ a positive function that vanishes exactly for $r > a$. For the sake of simplicity, we restrict the discussion to $s$-waves and the units employed are $\hbar=2m=1$, $m$ being the mass of the decaying particle.
As a consequence, the energy of the particle is denoted by $E=k^2$, with $k$ the corresponding wave number and hence the Hamiltonian to the system reads
\begin{equation}
H=- \frac{d^2}{dr^2}-iW(r),
\label{i1}
\end{equation}
As is well known, the time evolved wave function  $\psi(r,t)$ may be written in terms of the retarded Green's function of the problem $g(r,r';t)$ and the initial state $\psi(r,0)$ as,
\begin{equation}
\psi(r,t)=\int_0^a g(r,r\,';t)\psi (r\,',0)\,dr\,'.
\label{73}
\end{equation}
Moreover, the retarded Green's function $g(r,r\,';t)$ may be written, using Laplace transform techniques, as \cite{gcp76,gc10,gc11}
\begin{equation}
g(r,r';t)=\frac{1}{ 2 \pi i} \int_{C_0} G^+(r,r\,';k) {\rm e}^{-i k^2t} \,2kdk,
\label{74}
\end{equation}
where $G^+(r,r\,';k)$ corresponds to the outgoing Green's function of the problem and the integration contour $C_0$ goes along the first quadrant of the complex $k$ plane as indicated in
Fig. \ref{fig1}.

Our approach exploits the analytical properties of $G^+(r,r\,';k)$ in the complex $k$ plane. There,  for finite range potentials, it is well known that this function possesses an infinite number of complex poles. For real potentials the complex poles are located on the lower half of the $k$ plane and as a consequence of time reversal invariance, they are distributed symmetrically with respect to the imaginary $k$ axis. Hence, for a given pole at $k_p=\alpha_p - i \beta_p$, with $\alpha_p$ and $\beta_p$ positive quantities, there corresponds a pole at $k_{-p}=-k_p^*$ \cite{newtonch12}.  Bound and antibound poles correspond, respectively,  to purely imaginary positive and negative values of $k$. However, for complex potentials, time reversal considerations no longer apply \cite{joffily73,chgc13}. For absorptive potentials, causality prevents that poles sit on the first quadrant of the $k$ plane, however, they might appear on the other quadrants.  We denote by $k_p$  the poles on the fourth quadrant and by $k_{-p}$ the poles that are located on the second and third quadrants of the $k$ plane. A somewhat surprising feature of absorptive potentials is that they do not support bound or antibound poles \cite{joffily73,chgc13}.

Resonant states, also known as quasinormal modes \cite{leung98,starinets09,konoplya11} follow from  the residues at the poles of the outgoing Green's function to the problem \cite{gcp76,gc92,gc10,gc11}, that is,
\begin{equation}
\rho_p(r,r\,')= \frac{u_p(r)u_p(r\,')}{2k_p \left \{\int_0^a u_p^2(r) dr + i\frac{u_p^2(a)}{2k_p}\right \} },
\label{74b}
\end{equation}
which allows to write a normalization condition for these states as
\begin{equation}
\int_0^a u_p^2(r) dr + i\frac{u_p^2(a)}{2k_p}=1.
\label{74c}
\end{equation}
It may be shown analytically, that both the normalization condition proposed by Zel'dovich \cite{zeldovich61} and that corresponding to the complex scaling method \cite{giraud03} coincide exactly with the condition given by Eq. (\ref{74c}) \cite{gcm14}.
Resonant states satisfy the Schr\"odinger equation of the problem with outgoing boundary conditions, namely, using Eq. (\ref{i1}) one may write
\begin{equation}
[k_p^2-H]u_p(r)=0
\label{i2}
\end{equation}
with $u_p(0)=0$ and $u^{'}_p(a)=ik_pu_p(a)$, the prime denoting the derivative with respect to the variable $r$. Notice that the energy eigenvalues $k_p^2$ appearing in Eq. (\ref{i2}), which follow from the vanishing of the coefficients of the incoming waves, correspond, as mentioned above, to the poles of the outgoing Green's function $G^+(r,r';k)$.

A convenient form to evaluate Eq. (\ref{74}) is to close the contour indicated in Fig. \ref{fig1},  where the path $C_L$ corresponds to a straight line $45^{\circ}$ off the real $k$ axis that passes through the origin, and apply Cauchy's theorem \cite{gcp76}.
Taking the limit of the semi-circles radii $C_R$ to infinity, and noting that the factor $\exp(-ik^2t)$ in the corresponding integrands guarantees that the contribution of these contours vanish in that limit,  allows to rewrite Eq. (\ref{74}), using Eqs. (\ref{74b}) and (\ref{74c}), as a sum over exponentially decaying terms plus an integral contribution along the path $C_L$. Moreover, by performing in the integral contribution the change of variable $k=\gamma z$, with $\gamma=\sqrt{-i}$, we may write the resulting expression for $g(r,r\,';t)$ as
\begin{eqnarray}
g(r,r\,';t)&=& \sum_{p=1}^{\infty} u_p(r)u_p(r')e^{-i k_p^2t} + \nonumber \\ [.3cm]
&& {1 \over \pi} \int_{-\infty}^{\infty} G^+(r,r\,';\gamma z) {\rm e}^{-z^2t} \,zdz.
\label{74a}
\end{eqnarray}
The poles $k_p=\alpha_p-i\beta_p$, that appear in the exponentially decaying sum in Eq. (\ref{74a}),
are all located on the fourth quadrant of the $k$ plane. They are named \textit{proper poles} and satisfy $\alpha_p > \beta_p$. The corresponding complex energies $k^2_p=E_p=\mathcal{E}_p-i\Gamma_p$ define the resonance positions $\mathcal{E}_p=(\alpha_p^2-\beta_p^2)$ and the decaying widths $\Gamma_p=4\alpha_p\beta_p$, that fulfill $\mathcal{E}_p > \Gamma_p$.

The outgoing Green function $G^+(r,r\,';k)$ may be expanded as an infinite sum over the full set of resonant states to the problem along the internal interaction region, $r <a$ and $r\,' <a$ \cite{more71,gc76,gareev81} and also when either $r$ or $r\,'$ are evaluated at $a$ but no when $r=r'=a$ \cite{gcb79,gc10,gc11}. We denote the above circumstance with the notation $(r,r')^\dag \leq a$.
The validity of the above expansion is independent of whether the potential is real or complex \cite{gcb79,wromo80}. Hence we may write,
\begin{equation}
G^+(r,r\,';k) = \sum_{p=-\infty}^{\infty}  \frac {u_p(r)u_p(r\,')}{2k_p(k-k_p)}; \quad (r,r')^\dag \leq a
\label{9x}
\end{equation}
The outgoing Green's function satisfies the equation
\begin{equation}
[k^2-H]G^+(r,r\,';k)=\delta(r-r')
\label{i3}
\end{equation}
with boundary conditions $G^+(0,r\,';k)=0$ and $G'^{+}(a,r\,';k)=ikG^+(a,r\,';k)$.
Substitution of (\ref{9x}) into (\ref{i3}) yields the closure relation \cite{gc10,gc11},
\begin{equation}
\frac{1}{2}\,\sum_{p=-\infty}^{\infty} u_p(r)u_p(r\,')=\delta(r-r\,'); \quad (r,r')^\dag \leq a
\label{9y}
\end{equation}
and the sum rule,
\begin{equation}
\sum_{p=-\infty}^{\infty} \frac{u_p(r)u_p(r\,')}{k_p}=0; \quad (r,r')^\dag \leq a.
\label{9yy}
\end{equation}
Moreover, using the identity $1/[2k_p(k-k_p)] =(1/2k)[1/k_p+ 1/(k-k_p)]$ into (\ref{9x}), in view of  (\ref{9yy}), allows to write the alternative expansion
\begin{equation}
G^+(r,r\,';k) = \frac{1}{2k}\,\sum_{p=-\infty}^{\infty}  \frac {u_p(r)u_p(r\,')}{k-k_p}; \quad
(r,r')^\dag \leq a.
\label{i4}
\end{equation}
Then substituting (\ref{i4}) into (\ref{i3}) leads to the additional sum rule
\begin{equation}
\sum_{p=-\infty}^{\infty} u_p(r)u_p(r\,')k_p =0; \quad (r,r')^\dag \leq a.
\label{9yyy}
\end{equation}

Substitution of Eq. (\ref{9x}) into the integral term of  Eq. (\ref{74a}) leads to the representation of the retarded Green's function $g(r,r\,';t)$ as a sum of exponentially decaying terms plus a linear combination of resonant states and Moshinsky functions, which becomes relevant at very short or very long times compared with the lifetime of the system \cite{gc10,gc11}. In particular, the long time behavior of $g(r,r\,';t)$ may be obtained by expanding the Moshinsky functions at long times,  which go as $a/(k_pt^{1/2}) +b/(k_pt^{3/2})+...$, with $a$ and $b$ constants. This requires using the sum rule given by Eq. (\ref{9yy}) to eliminate the $t^{-1/2}$ contribution to yield the well known $t^{-3/2}$ asymptotic long time behavior. Clearly this may become cumbersome in numerical calculations, since either one has to take out explicitly the term $t^{-1/2}$ in the above expansions or to perform the calculations with a very large number of poles.

Here we follow a similar approach to that of Ref. \cite{gcmv07}, which addressed real potentials, to obtain the behavior of $g(r,r\,';t)$ at long times. This procedure does not rely on using the above sum rules and leads directly to the correct long-time asymptotic behavior.  It exploits the fact that at long times compared with the lifetime of the system $\tau$, the integrand over the integral term in Eq. (\ref{74a}) oscillates widely and hence it may be evaluated, to a very good approximation,  by the steepest descent method \cite{erdelyi}. One sees that the saddle point of the exponential in Eq. (\ref{74a}) is at $z=0$ and hence one may perform a Taylor expansion of $G^+(r,r';\gamma z)$ around that value, namely,
\begin{equation}
G^+(r,r';\gamma z)=G^+(r,r\,';0) + \gamma z\left [\frac{\partial}{\partial \gamma z} G^+(r,r\,';\gamma z)\right]_{z=0}+.. \label{2}
\end{equation}
Substitution of Eq.\ (\ref{2}) into the integral term in Eq. (\ref{74a}) leads to an expression where one sees that the term proportional to $G^+(r,r\,';0)$ vanishes exactly because the integrand is an odd function of $z$. The $z$-integral for the next term in the Taylor expansion may be evaluated by making the change of variable  $u=zt^{1/2}$, which gives the leading term as the inverse power in time $t^{-3/2}$. Consequently, at long times compared with the lifetime $\tau$ of the system, Eq. (\ref{74a}) may be written approximately as,
\begin{eqnarray}
&&g(r,r\,';t) \approx \sum_{p=1}^{\infty} u_p(r)u_p(r')e^{-i\mathcal{E}_pt}e^{-\Gamma_pt/2} + \nonumber \\ [.3cm]
&&\eta \left \{ \frac{\partial}{\partial k} G^+(r,r';k)\right\}_{k=0}
\frac{1}{t^{3/2}};\quad (r,r')^\dag \leq a
\label{3}
\end{eqnarray}
where $\eta=1/(4\pi i)^{1/2}$ and the exponentially decaying terms are written explicitly in terms of the \textit{proper poles} $\mathcal{E}_p$ and $\Gamma_p$ defined previously. In general it is difficult to obtain a closed analytical expression for the factor $[\partial G^+(r,r\,';k)/\partial k]_{k=0}$. One may use however, the expansion of the outgoing Green function given by Eq. (\ref{9x}) to evaluate this factor and write Eq. (\ref{3}) as,
\begin{eqnarray}
&& g(r,r\,';t) \approx \sum_{p=1}^{\infty} u_p(r)u_p(r\,')e^{-i\mathcal{E}_pt}e^{-\Gamma_pt/2} - \nonumber \\ [.3cm]
&& \eta\, \sum_{p=-\infty}^{\infty}\frac{u_p(r)u_p(r\,')}{2k_p^3} \frac{1}{t^{\,3/2}};\quad (r,r')^\dag \leq a.
\label{3a}
\end{eqnarray}
\subsection{Decaying wave function and survival probability}\label{subdos}

Then, inserting Eq. (\ref{3a}) into Eq. (\ref{73}) yields the expression for the time dependent wave function,
\begin{eqnarray}
\psi(r,t) \approx \sum_{p=1}^{\infty} C_pu_p(r)e^{-i\mathcal{E}_pt}e^{-\Gamma_pt/2} - \nonumber \\ [.3cm]
\eta\, \left\{\sum_{p=-\infty}^{\infty}\frac{C_pu_p(r)}{2k_p^3} \right\} \,\frac{1}{t^{\,3/2}};\,\,r \leq a,
\label{3b}
\end{eqnarray}
where the overlap coefficients $C_p$ are defined as,
\begin{equation}
C_p=\int_0^a \psi(r,0) u_p(r) dr.
\label{3c}
\end{equation}
Assuming that the initial state $\psi(r,0)$ is normalized to unity, we may use the closure relation given by Eq. (\ref{9y}) to write,
\begin{equation}
\frac{1}{2} \sum_{p=1}^{\infty}\left \{  C_p \bar{C}_p  + C_{-p} \bar{C}_{-p}\right \}= 1,
\label{9z}
\end{equation}
where
\begin{equation}
{\bar C}_p=\int_0^a \psi^*(r,0) u_p(r) dr.
\label{3e}
\end{equation}

Equation (\ref{3b}) provides the time evolution of the decaying wave function for an absorptive potential as an expansion in terms of resonant states along the exponentially decaying and long-time regimes.

The survival amplitude gives the probability amplitude that at time $t$ the decaying particle remains in the initial state,
\begin{equation}
A(t) = \int_0^a \psi^*(r,0)\psi(r,t) \,dr.
\label{90}
\end{equation}
Hence, the corresponding survival probability is given by the expression,
\begin{equation}
S(t)=|A(t)|^2.
\label{3e}
\end{equation}

Substitution of Eq. (\ref{3b}) into Eq. (\ref{90}) yields
\begin{eqnarray}
A(t) &\approx & \sum_{p=1}^{\infty} C_p{\bar C}_pe^{-i\mathcal{E}_pt}e^{-\Gamma_pt/2} -\nonumber \\ [.3cm]
&& \eta\, \sum_{p=1}^{\infty}\left \{\frac{C_p{\bar C}_p}{2k_p^3} + \frac{C_{-p}{\bar C}_{-p}}{2k_{-p}^3}\right \} \frac{1}{t^{\,3/2}}.
\label{3d}
\end{eqnarray}

Equation (\ref{3d})  has some resemblance to an expression of the survival amplitude derived by Longhi for a non-Hermitian Friedrichs-Fano-Anderson model \cite{longhi09}. Our approach, however  provides an explicit resonant expansion for the nonexponential contribution.

It is worth recalling that for real potentials
\begin{equation}
k_{-p}=-k^*_p,   \qquad u_{-p}(r)=u^*_p(r).
\label{3pa}
\end{equation}
and as  a consequence, the survival amplitude corresponding to real potentials reads \cite{gcmv07},
\begin{equation}
A(t) \approx \sum_{p=1}^{\infty} C_p{\bar C}_pe^{-i\mathcal{E}_pt}e^{-\Gamma_pt/2} -i \eta\,  {\rm Im}
\sum_{p=1}^{\infty}\left \{\frac{C_p{\bar C}_p}{k_p^3} \right \} \frac{1}{t^{\,3/2}}.
\label{3dr}
\end{equation}
Similarly, for real potentials, since in view of (\ref{3pa}), $C_{-p}{\bar C}_{-p}=(C_p{\bar C}_p)^*$,      Eq. (\ref{9z}) becomes,
\begin{equation}
{\rm Re}\, \sum_{p=1}^{\infty}\left \{C_p \bar{C}_p\right \}= 1.
\label{9zr}
\end{equation}

One may associate to each decaying width $\Gamma_p$, a time scale $\tau_p=1/\Gamma_p$. The longest of these times scales, which corresponds to the shortest decaying  width, defines the lifetime $\tau$ of the system. For single particle potentials, the shortest decaying width is usually that with the lowest resonance energy, this is for $p=1$, and hence,
\begin{equation}
\tau=\frac{1}{\Gamma_1}.
\label{tau}
\end{equation}

As a result of the above considerations, after a few lifetimes, using Eq. (\ref{3d}) into the definition of the survival probability $S(t)$ given by Eq. (\ref{3e}), one sees that the only term that survives in the sum of the exponentially decaying terms is that corresponding to the longest lifetime, i.e., $S(t) \approx |C_1{\bar C}_1|^2\exp(-\Gamma_1t)$.  As time evolves, that exponentially decaying term becomes of the same order of magnitude as the corresponding nonexponential $t^{-3}$ long-time contribution of $S(t)$.
The time $t_{tr}$ at which the exponential-nonexponential transition occurs depends on each specific system. Clearly, a similar argument holds for the case of real potentials. Here, it has been found that the exponential-nonexponential long-time transition depends on the value of the quantity  $R=\mathcal{E}_r/\Gamma_r$ \cite{gcrr01,gcv06}. Commonly, since the poles are proper, $R > 1$, and hence the above transition occurs after a number of lifetimes. In the numerical example discussed below the transition occurs around $t_{tr} \approx 27 \tau$ as shown in Fig. \ref{fig4}. For times smaller than the transition time $t_{tr}$, the time inverse power $t^{-3}$ contribution is much smaller than the exponential ones due to the multiplying  factors that go as $1/k_r^6$, with $r=\pm p$, that appear in the former contribution, as exemplified in Fig. \ref{fig5}.

A point worth emphasizing refers to Eqs. (\ref{9z}) and (\ref{9zr}), which follow, respectively for complex and real potentials, from the closure relationship (\ref{9y}). The coefficients $C_p$ cannot be interpreted as probability amplitudes, since the sum of the their square moduli does not add up to the norm of the expanded function. However, although $C_p \bar{C}_p$ might be negative, for real potentials its real part adds up to unity, and if the `strength' value of, say, $C_r \bar{C}_r \approx 1$, then that term dominates the expansion of the exponential terms of the survival amplitude. A more involved situation may occur for the terms  $[C_r \bar{C}_r  + C_{-r} \bar{C}_{-r}]/2$ appearing in (\ref{9z}), as discussed in the  model calculation of the next section.

We illustrate in the next Section, that by choosing an initial state which overlaps strongly with a state corresponding to a pole located on the second or third quadrants of the $k$ plane, specifically with a \textit{spectral singularity resonant state} \cite{chgc13}, may lead to a modification of the values of the coefficients corresponding to the proper poles located on the fourth quadrant. This in turn, leads to interference contributions that change the exponential decaying behavior of the survival probability into an  oscillating contribution at time scales of the order of a lifetime  of the system. Such a behavior is a consequence of the lack of time-reversal symmetry in the position of the poles $k_p$ and $k_{-p}$ for absorptive potentials. It involves  consideration only of the exponentially decaying terms appearing in Eq. (\ref{3d}), since at that time scales the nonexponential $t^{-3/2}$ contribution is negligible, as pointed out previously.

\section{Model}\label{tres}

We consider a purely absorptive delta-shell potential of radius $a$. Then, in view of (\ref{ia1}), we may write
\begin{equation}
V(r)=-ib\delta(r-a),
\label{a11}
\end{equation}
with $b > 0$. As discussed in detail in Ref. (\cite{chgc13}), the solutions to Eq. (\ref{i2}) obeying outgoing boundary conditions read,
\begin{equation}
u_p(r)=A_p \,\sin (k_pr), \quad r\leq a
\label{e2i}\
\end{equation}
and
\begin{equation}
u_p(r)=B_p \,e^{ik_pr}, \quad r \geq a,
\label{e2e}
\end{equation}
from which one readily obtains, using the continuity of the solutions at $r=a$ and  the discontinuity of the corresponding derivatives due to the $\delta$ interaction, the equation for the poles
\begin{equation}
2k_p -b\, (e^{2ik_p a}-1)=0.
\label{e3}
\end{equation}

\begin{figure}[!tbp]
\begin{center}
\includegraphics[width=3.7in]{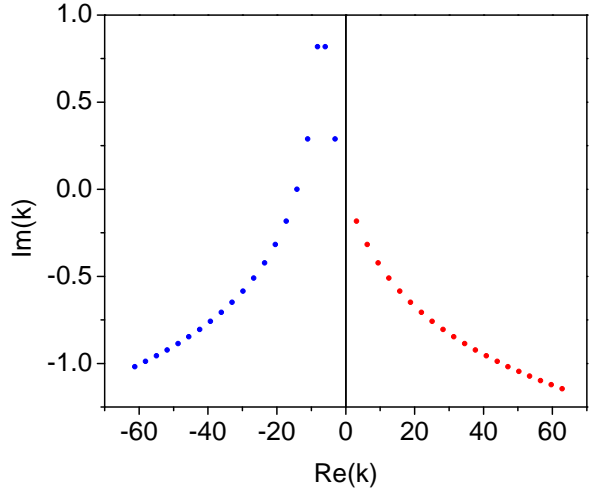} \caption{ \footnotesize (color online) Distribution of poles for an intensity of the $\delta$-shell potential $b=9\pi/2$ of radius  $a=1$. }
\label{fig2}
\end{center}
\end{figure}

The solution to Eq. (\ref{e3}) yields the set of poles  to the problem as a function of the potential parameters \cite{chgc13}. Figure \ref{fig2} exhibits the distribution of the first $40$ poles for $b=9\pi/2$ and $a=1$. We have chosen that  value of the intensity to guarantee that there is a spectral singularity located precisely at $k_{-5}=-9\pi/2$. Notice, as pointed out in Sec. \ref{dos}, and shown  explicitly in Table \ref{tablap}, that the absence of time reversal invariance implies that the poles $\{k_{-p}\}$  sitting on the second and third quadrants of the $k$ plane are non symmetrical with respect to those located on the fourth quadrant. Moreover the real part of the poles $k_{-p}$ located in the  third quadrant of the $k$ plane tend to values proportional to $\pi/2$ as $p$ increases, whereas the real part of the poles on the second quadrant tend to be proportional to $\pi$ as $p$ decreases. On the other hand, the real parts of the poles located on the fourth quadrant are proportional to $\pi$.
\begin{table}[!tbp]
\caption{\label{tablap} Poles  $k_{-p}$ and $k_{p}$ for the imaginary $\delta$-shell potential
with the same parameters as in Fig. \ref{fig2}.}
%\renewcommand{\arraystretch}{1.3}
%\centering
\begin{ruledtabular}
\begin{tabular}{ccccc}
%   \hline
   $n$ & $\alpha_{-p}$  &  $\beta_{-p}$ & $\alpha_{p}$  &  $\beta_{p}$\\
%    \hline
    \hline
   1 & -3.10532&	0.28798&	3.13260	&-0.18350\\
%    \hline
  2 & -5.96420&	0.81871&	6.27128&	-0.31770\\
%  \hline
  3& -8.17293&	0.81868&	9.41193&	-0.42343\\
%  \hline
  4& -11.03184&	0.28796	&12.55336&	-0.51067\\
%  \hline
  5&  -14.13717&	0.00001&	15.69512&	-0.58492\\
%  \hline
  6& -17.26976&	-0.18351&	18.83702&	-0.64956\\
%  \hline
  7& -20.40845&	-0.31770	&21.97898&	-0.70679\\
%  \hline
  8& -23.54910&	-0.42343&	25.12097&	-0.75813\\
%  \hline
  9& -26.69053&	-0.51067&	28.26295&	-0.80469\\
%  \hline
  10& -29.83228&	-0.58493&	31.40492&	-0.84728\\
%    \hline
\end{tabular}
\end{ruledtabular}
\end{table}
Using the solution for $u_p(r)$ along the internal region $r \leq a$ given by (\ref{e2i}) allows to obtain, using (\ref{74c}), the expression for normalization coefficient,
\begin{equation}
A_p=\left [ \frac{2(-iba-2i k_p a)}{a(1-iba-2ik_p a)} \right ]^{1/2}.
\label{a7}
\end{equation}
This allows to obtain the full set of resonant solutions $\{u_p(r)\}$ along the internal region $r \leq a$ which are necessary to evaluate the expansion coefficients $C_p$ and ${\bar C}_p$, given respectively by, (\ref{3c}) and (\ref{3e}). In order to calculate these coefficients one needs to specify the initial state $\psi(r,0)$.

A frequent choice is an initial state that may overlap strongly with a given resonant state of the system.
In this sense, an appropriate choice is the wave function along the internal interaction region,
\begin{equation}
\psi(r,0)= N_c \sin (k_cr), \qquad r \leq a
\label{a4}
\end{equation}
with $k_c$ is an arbitrary real parameter. Imposing the normalization condition
\begin{equation}
\int_0^a |\psi(r,0)|^2 dr=1
\label{a4.1}
\end{equation}
one obtains \cite{gcmm96},
\begin{equation}
N_c =\frac{ \sqrt{2/a} }{[ 1-(2k_c a)^{-1} \sin (2k_ca)]^{1/2}}.
\label{a5}
\end{equation}
The initial state given by (\ref{a4}) represents the decaying portion of a state that in addition possesses an external portion that initially is outside from the interaction region and therefore will not be considered here. Using Eqs. (\ref{e2e}), (\ref{a7}), (\ref{a4}) and (\ref{a5}) into (\ref{3c}) yields,
\begin{equation}
C_p=N_cA_p \frac{-k_p \sin (k_ca) \cos (k_pa) + k_c \sin (k_pa) \cos (k_ca)}{k_p^2 -k_c^2}.
\label{a8}
\end{equation}
\begin{figure}[!tbp]
\begin{center}
\includegraphics[width=3.2in]{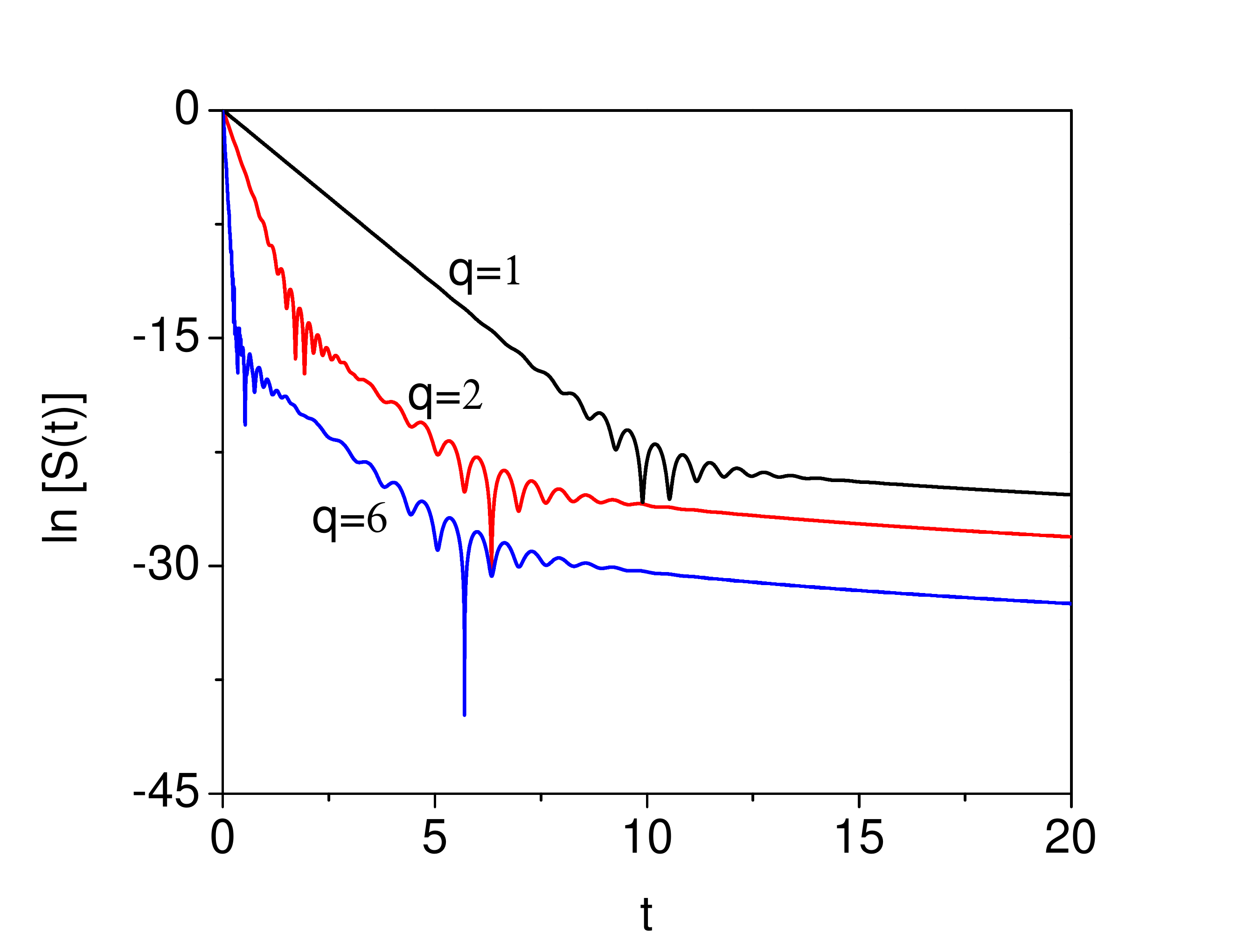}
\caption{ \footnotesize (color online) Plot of ln [S(t)] as a function of the time $t$ for the imaginary $\delta$-shell potential with parameters $b=9\pi/2$ and $a=1$ and initial infinite box states with $q=1,2,6$ which exhibit, respectively, maximum overlap with the resonant states with $p=1,2,6$, which correspond to poles sitting on the fourth quadrant of the $k$ plane. See text.}
\label{fig3}
\end{center}
\end{figure}
\begin{figure}[!b]
\begin{center}
\includegraphics[width=3.2in]{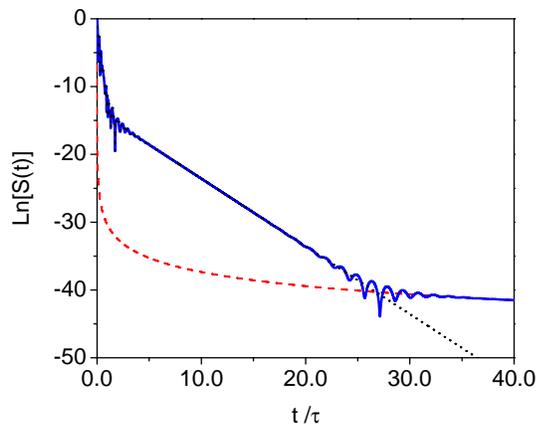}
\caption{ \footnotesize (color online) Plot of the ln [S(t)] as a function of the time $t$ in units of the lifetime $\tau=1/\Gamma_1$ for the purely imaginary $\delta$-shell potential with parameters $b=9\pi/2$ and $a=1$ for the initial state given by Eq. (\ref{a4}) with $k_c=9\pi/2$, which exhibits first a transition to the lowest decaying state with decaying width $\Gamma_1$ (solid line), that arises, respectively, from the exponential decaying contribution (dot line) followed by a transition to an inverse power $t^{-3}$ behavior at long times (dash line). See text.}
\label{fig4}
\end{center}
\end{figure}

A common choice for the initial state is the infinite box model state, which requires  to have $k_c=q\pi/a$ with $q=1,2,...$ and hence, in view of (\ref{a5}), it yields the well known normalization $N_c=\sqrt{2/a}$ \cite{winter61,gcmv07}. Using the potential parameters $b=9\pi/2$ and $a=1$, one may readily obtain the corresponding survival probabilities $S(t)$ using Eqs. (\ref{3d}) and (\ref{3e}). This is shown in Fig. \ref{fig3} for  $q=1,2,6$. For these choices the decay is initially dominated, respectively, by the exponential decay widths $\Gamma_1$, $\Gamma_2$ and $\Gamma_{6}$. This follows because the corresponding expansion coefficients in (\ref{3d}) are dominated, in view of Eq. (\ref{9z}), respectively by ${\rm Re}\{C_1^2\} =1.0129$, ${\rm Re}\{C_2^2\} =1.0355$ and ${\rm Re}\{C_{6}^2\}=1.1494$, all of them of the order of unity. The above quantities follow because there is a maximum overlap with the corresponding resonant states defined by Eq. (\ref{e2i}) with $p=1,2,6$. As shown in Fig. \ref{fig3}, eventually, the cases with $q=2$ and $q=6$  suffer a transition to the state with the smallest width, $\Gamma_1$, and finally, all cases exhibit a transition to the inverse power long time behavior as $t^{-3}$ as follows from Eq. (\ref{3d}). In this case $\mathcal{E}_1=9.7795$ and $\Gamma_1=2.2993$ and hence $R=\mathcal{E}_1/\Gamma_1=4.25$ and the exponential-nonexponential transition time $t_{tr}$ occurs at roughly $t_{tr} \approx 27 \tau$.

\begin{figure}[!tbp]
\begin{center}
\includegraphics[width=3.2in]{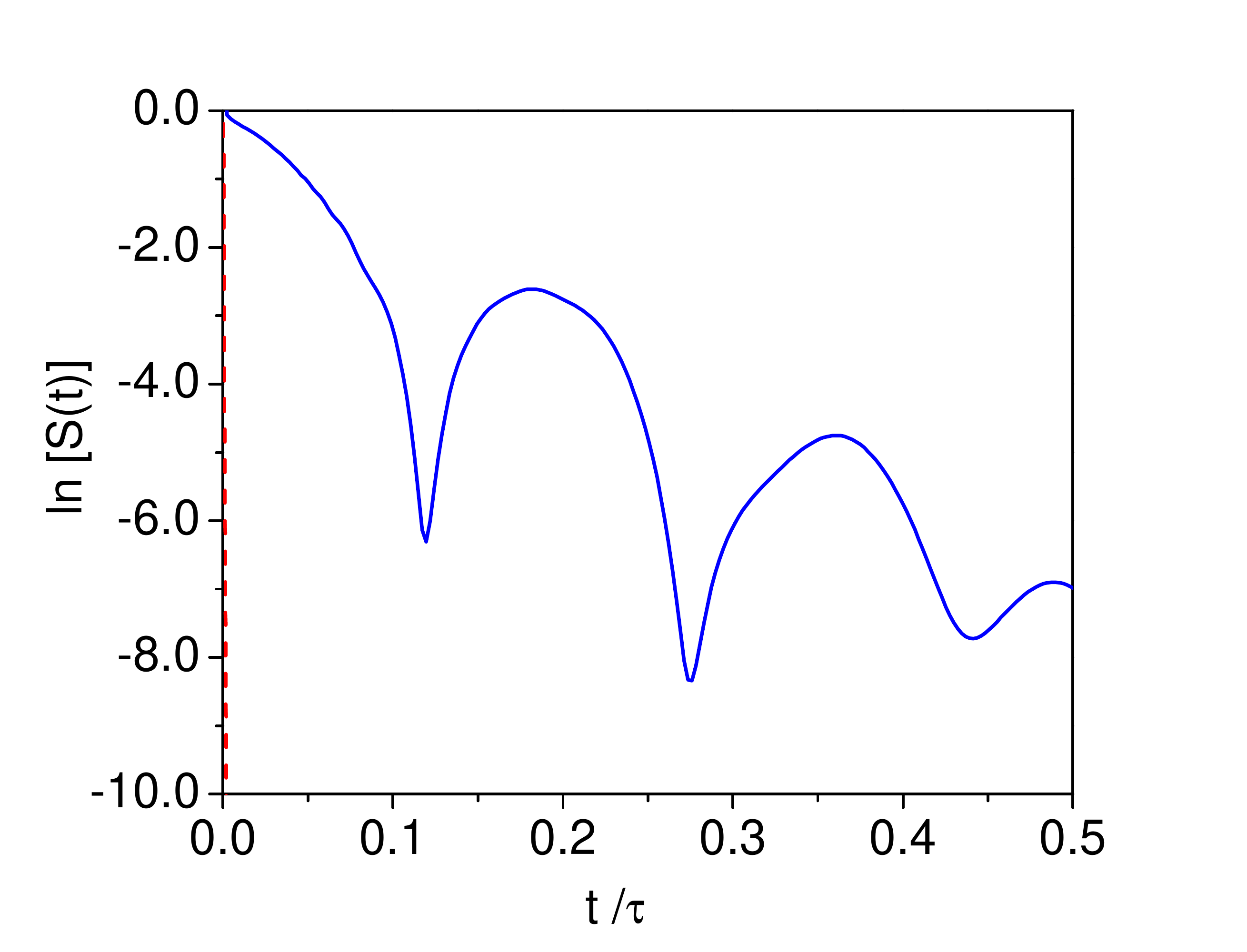}
\caption{ \footnotesize (color online) Plot of the ln [S(t)] as a function of the time $t$ in units of the lifetime $\tau=1/\Gamma_1$ for the purely imaginary $\delta$-shell potential considered in Fig. \ref{fig4}, that exhibits an oscillatory behavior at short times which arises as a consequence of the strong overlap of the initial state with a \textit{spectral singularity resonant function} (solid line) and a negligible inverse power of time contribution (dashed line). See text.}
\label{fig5}
\end{center}
\end{figure}

It is worth mentioning that the above behavior of the survival probability for the \textit{imaginary potential} with the choice of an initial state having maximum overlap with states corresponding to poles located on the fourth quadrant of the $k$ plane is qualitatively similar to that of a  \textit{real potential}, namely, it exhibits an exponential decaying regime followed by a nonexponential contribution at long times \cite{gcmv07}.

\begin{table}
\caption{\label{tab} Coefficients  $C_{-p}^2$ and $C_{p}^2$ for the imaginary $\delta$-shell potential
with the same  parameters as in Fig. \ref{fig4} and an initial state with $k_c=9\pi/2$. See text.}
%\renewcommand{\arraystretch}{1.3}
%\centering
\begin{ruledtabular}
\begin{tabular}{ccccc}
%   \hline
   $p$ & $Re(C_{-p}^2)$  &  $Im(C_{-p}^2)$ & $Re(C_p^2)$ & $Im(C_p^2)$\\
%    \hline
    \hline
   1 &  0.00108 & -0.00039 & 0.00111 &	-0.00020\\
%    \hline
  2 &   0.00357 &  -0.00644 &	0.00662 &	-0.00131\\
%  \hline
  3&  -0.00817 &	 -0.00757 &	0.03260  &	-0.00914\\
%  \hline
  4&  -0.00663  &  -0.00112 &	\textbf{0.31311}   &	-0.26619\\
%  \hline
  5&    {\bf 0.99502}	  &   0.07038	 & \textbf{0.44151}   &     0.33393\\
%  \hline
  6&  -0.00419 &	-0.00017	 &  0.08426  &	 0.01809\\
%  \hline
  7&  -0.00370 &	-0.00001	 &  0.03770  &	 0.00464\\
%  \hline
  8&  -0.00335 &	-0.00006	 &  0.02282  &	 0.00195\\
%  \hline
  9&  -0.00308 &	-0.00004	 &  0.01595  &	 0.00105\\
%  \hline
  10&  -0.00286 &	-0.00002	 &  0.01212  &	 0.00064\\
%    \hline
\end{tabular}
\end{ruledtabular}
\end{table}
However, the choice of an initial state with maximum overlap with a resonant state corresponding to a pole on the second or third quadrants of the $k$ plane, in particular with a \textit{spectral singularity}, leads to a novel effect which follows from the lack of time reversal symmetry of the distribution of poles.

Let us illustrate this by considering the imaginary $\delta$-shell potential with the same parameters as in the previous case, namely, $b=9\pi/2$ and $a=1$. This yields a \textit{spectral singularity} pole located at $k_{-5}=-9\pi/2$, as shown in Table \ref{tablap}.   Figure \ref{fig4} exhibits a plot of $\ln [S(t)]$ as  a function of the time $t$. Apparently, the behavior of the survival probability looks similar to the cases displayed in Fig. \ref{fig3}. However, on close scrutiny at  times of the order of a lifetime, where the exponential contribution dominates, namely, the survival amplitude is given by the first term on the right-hand side of Eq. (\ref{3d}), one obtains the unexpected behavior shown  in Fig. \ref{fig5}, which corresponds to a nonexponential oscillatory dependence with time. This behavior may be understood by inspection of Table \ref{tab}. One sees  that a maximum overlap is exhibited by the coefficient  ${\rm Re}\,C_{-5}^2=0.99502$.  This favors a large overlap for the coefficients ${\rm Re}\,C_{4}^2=0.31311$ and ${\rm Re}\,C_{5}^2=0.44151$. In fact these three coefficients almost satisfy the relationship (\ref{9z}). This means that the initial state with $k_c=9\pi/2$, which overlaps strongly with the state $u_{-5}(r)$, shares also its strength with the states $u_4(r)$ and $u_5(r)$. The reason is that the value $k_c=9\pi/2$ of the initial state is approximately half-between $\alpha_4$ and $\alpha_5$, as follows by inspection of Table \ref{tablap}. As a consequence the survival amplitude given by Eq. (\ref{3d}) may be written approximately as,
\begin{equation}
A(t) \approx   C_4^2e^{-i\mathcal{E}_4t}e^{-\Gamma_4t/2} + C_5^2e^{-i\mathcal{E}_5t}e^{-\Gamma_5t/2},
\label{3da}
\end{equation}
and hence, one sees that the interference contribution to the corresponding survival probability
exhibits an oscillatory behavior that disappears gradually as time evolves. An accurate description, requires however, to take several pole terms into account. It is worth mentioning that for \textit{real potentials}, time reversal invariance implies, as pointed out above, that $C_{-p}{\bar C}_{-p}=
(C_p{\bar C}_{p})^*$, and therefore the above situation does not occur, except when a set of complex poles lie very close together as in one-dimensional multibarrier resonant tunneling systems \cite{gcrv07,gcrv09}.

\section{concluding remarks}\label{cuatro}

The main result of this work is to show that the time evolution of decay for absorptive imaginary potentials may lead, due to the lack of time reversal invariance, to an unexpected behavior that originates in the choice of the initial state. We find for an exactly solvable model, that if the initial state overlaps strongly with a state corresponding to a pole on the fourth quadrant, then the survival probability is not qualitatively different from that of a real potential. However, if the initial state overlaps strongly with a resonant state corresponding to poles located on the second or third quadrants of the $k$ plane, such that $\alpha_{-p} \neq \alpha_p$, as for a \textit{spectral singularity}, then the survival probability may exhibit instead of exponential decay, at time scales of the order of a lifetime of the system, a nonexponential almost harmonic oscillatory behavior.

In a recent paper \cite{gcmv13}, the probability density $|\psi(r,t)|^2$ for propagation along the external interaction region at asymptotic long distances and times was considered for real potentials. This refers to an unexplored postexponential regime that provides the ultimate fate of a decaying particle. We believe that it would be of interest to investigate that regime for  purely absorptive potentials to find out if the oscillatory behavior in time found here for the survival probability prevails.

We end by commenting that would be interesting to look for the possibility of considering the lack of time reversal symmetry to design quantum systems that would exhibit the time evolution of decay described here. We hope the present work may stimulate further work on this subject.

\section*{acknowledgments}
G. Garc\'ia-Calder\'on acknowledges partial financial support form DGAPA-UNAM under grant IN111814.
\end{document}